\documentclass[aps,prl,twocolumn,superscriptaddress,showpacs]{revtex4}

\usepackage{epsfig}
\usepackage{dcolumn}
\usepackage{bm}

\begin{document}

\title{Intrinsic optical dichroism in the chiral superconducting state of 
Sr$_{2}$RuO$_{4}$ }
\author{K. I. Wysoki\'nski}
\affiliation{Institute of Physics, M. Curie-Sk\l {}odowska University, 
Radziszewskiego
10, PL-20-031 Lublin, Poland}
\affiliation{Max Planck Institut f\"ur Physik komplexer Systeme, 
D-01187 Dresden, Germany}
\author{James F. Annett}
\affiliation{H. H. Wills Physics Laboratory, University of Bristol, 
Tyndall Ave, BS8-1TL,
UK.}
\author{B. L. Gy\"orffy}
\affiliation{H. H. Wills Physics Laboratory, University of Bristol, 
Tyndall Ave, BS8-1TL,
UK.}
\date{ \today }

\begin{abstract}
We present an analysis of the Hall conductivity $\sigma _{xy}(\omega ,T)$ in
time reversal symmetry breaking states of exotic superconductors. We find
that the dichroic signal is non-zero in systems with inter-band order
parameters. This new intrinsic
mechanism may explain the Kerr effect observed in strontium ruthenate and
possibly other superconductors. We predict coherence factor effects in the
temperature dependence of the imaginary part of the ac Hall conductivity $
Im\sigma _{xy}(\omega ,T)$, which can be tested experimentally.
\end{abstract}

\maketitle

%
%

A superconducting state with chiral p-wave symmetry is of general
interest because it is the charged analouge of the superfluid $A$ phase of 
$^{3}He$.  In such a state the Cooper pairs are spin triplets and have a relative
angular momentum $l=1$, and therefore it occupies a unique place on the list
of superfluid phases of matter. Furthermore, this particular state has been
identified recently as a possible topological superconductor emphasizing
its relavence to understanding of superfluidity at the deepest level 
\cite{QZ10}. The best candidate to host this exotic state of matter appears to
be $Sr_{2}RuO_{4}$ \cite{Mackenzie2003}. The central supporting
evidence for chiral p-wave symmetry in this material
are experiments which show that the
superconducting state breaks time reversal 
invariance \cite{Luke1998,Kapitulnik2009}.

The possibility of using optical dichroism to detect time reversal symmetry
breaking (TRSB) pairing states in unconventional superconductors was first
suggested in the late 1980's \cite{Klemm1988,Joynt1991,Hirschfeld1989,Yip1992}.
Recently, such
dichroism was observed in polar Kerr effect measurements of the $1.5$K
superconductor Sr$_{2}$RuO$_{4}$ by Xia 
\textit{et al.} \cite{Xia2006}. Subsequently similar dichroism was
found in some underdoped high temperature superconductors \cite{Kapitulnik2009}.
The measurements on strontium 
ruthenate showed a small Kerr rotation of light of wavelength $\lambda =1550$
nm, corresponding to a rotation of the plane of polarization by an amount
approaching $100$ nrad at $T=0$ and going to zero at $T_{c}$ approximately
linearly in $T_{c}-T$. 
Strong evidence for TRSB in strontium ruthenate had
previously been seen in muon spin rotation \cite{Luke1998}, where the signal
shows a broadly similar temperature dependence. Together these observations
support the identification of this material as a chiral p-wave
superconductor \cite{Mackenzie2003}. However, the theoretical interpretation
of both of these experiments is difficult and edge currents predicted by the
chiral pairing theory have not been observed \cite{Hicks2010,Ashby2009}
leaving the question of the identification of the pairing state partially
unresolved \cite{Kallin2008}.

In particular the origin of dichroism in a chiral superconducting
state has attracted considerable attention in the recent literature 
\cite{Yakovenko2007,Mineev2008,Roy2008,Lutchyn2008,Goryo2008,Lutchyn2009,Goryo2010}. 
The conclusion of this work is that the dichroic signal is exactly zero in
the intrinsic limit, and only appears as a higher order effect in the
presence of impurity scattering \cite{Goryo2008,Lutchyn2009,Goryo2010}.
Numerical estimates of the Kerr signal arising from this mechanism appear
consistent with the experimental observations \cite{Kapitulnik2009}.

In this letter we propose a different mechanism for the generation of the
dichroic signal, which is purely intrinsic and does not rely on impurity
scattering or a finite width of the incident photon beam. The
principal difference between this work and the earlier calculations is that
our theory is based upon a multi-band pairing model of Sr$_{2}$RuO$_{4}$,
and, as we show below, the dichroic signal arises from inter-orbital pairing
associated with the $d_{xz}$ and $d_{yz}$ Ru orbitals. We have previously
shown that this same model gives a good description of the overall
thermodynamic properties of Sr$_{2}$RuO$_{4}$ \cite{Annett2002,Annett2003,Annett2006}. 
Crucially, the same inter-orbital pairing
model predicts a finite orbital magnetic moment on each Ru atom \cite{Annett2009},
which has the same origin as the calculated dichroic signal.
The two are in fact directly linked by the f-sum rule \cite{Souza2008}. The
fact that inter-orbital pairing associated with the Ru $d_{xz}$ and $d_{yz}$
is the key physical feature of dichroism in this theory is qualitatively
consistent with the proposals by Raghu, Kapitulnik and Kivelson \cite{Raghu2010}, 
however in our phenomenological theory all bands are assumed to
be superconducting with comparable values of the gap \cite{Annett2002}.

Our calculation of the optical dichroism is based on the systematic analysis
of the Bogoliubov de Gennes (BdG) equations developed by Capelle, Gross and
Gyorffy \cite{Capelle1997,Capelle1998}. They disscuss a
fairly complete list of conditions, including TRSB, under which dichroism
in the electromagnetic response of a superconductor occurs. In this
formalism the conductivity tensor can be expressed in terms of the
electromagnetic power absorption ${P}(\omega ,\bm{\epsilon})$ for light of
left and right circular polarizations, $\bm{\epsilon}_{L}$ and $\bm{\epsilon}_{R}$,
 respectively, 
\begin{equation}
\mathrm{Im}[\sigma _{xy}(\omega )]=\frac{1}{VE_{0}^{2}}\left[ {P}(\omega ,
\bm{\epsilon}_{L})-{P}(\omega ,\bm{\epsilon}_{R})\right] .  \label{imsxy}
\end{equation}
Here $V$ is the sample volume, $E_{0}$ is the electric field strength of the
light, and $\bm{\epsilon}_{L/R}=(1,\pm i,0)/\sqrt{2}$. Within the BdG
formalism the absorption spectrum can be calculated directly in terms of the
dipole matrix elements \cite{Capelle1997,Capelle1998} 
\begin{eqnarray}
{P}(\omega ,\bm{\epsilon}) &=&\frac{\pi ^{2}e^{2}E_{0}^{2}}{2\omega }
\sum_{N,N^{\prime },\mathbf{k}}f(E_{N}(\mathbf{k}))[1-f(E_{N^{\prime }}(
\mathbf{k}))]  \label{eq:p} \\ 
&\times& \mid \langle N^{\prime }\mathbf{k}\mid \hat{H}_{I}(\bm{\epsilon}
)\mid N\mathbf{k}\rangle \mid ^{2}\delta (E_{N^{\prime }}(\mathbf{k})-E_{N}(
\mathbf{k})-\hbar \omega), \nonumber 
\end{eqnarray}
where 
\begin{equation}
\mid N\mathbf{k}\rangle =\left( 
\begin{array}{c}
u_{N}(\mathbf{k}) \\ 
v_{N}(\mathbf{k})
\end{array}
\right)
\end{equation}
is the $N$th eigenvector of the BdG equation at wave vector $\mathbf{k}$ 
fulfilling the equation 
\begin{equation}
\left( 
\begin{array}{cc}
\hat{H}_{0}(\mathbf{k}) & \hat{\Delta}(\mathbf{k}) \\ 
\hat{\Delta}(\mathbf{k})^{\dagger } & -\hat{H}_{0}(\mathbf{k})^{\ast }
\end{array}
\right) \left( 
\begin{array}{c}
u_{N}(\mathbf{k}) \\ 
v_{N}(\mathbf{k})
\end{array}
\right) =E_{N}\left( 
\begin{array}{c}
u_{N}(\mathbf{k}) \\ 
v_{N}(\mathbf{k})
\end{array}
\right) .
\end{equation}
Here $\hat{H}_{0}(\mathbf{k})$ is the normal state tight binding
Hamiltonian, $\hat{\Delta}(\mathbf{k})$ is the matrix of gap parameters in
the tight binding spin-orbital basis. The matrix elements of the
light-matter interaction Hamiltonian in Eq.~(\ref{eq:p}) have the general form 
\begin{eqnarray}
\langle N^{\prime }\mathbf{k}\mid \hat{H}_{I}\mid N\mathbf{k}\rangle &&= 
\nonumber \\
\left( u_{N^{\prime }}^{\ast }(\mathbf{k}),v_{N^{\prime }}^{\ast }(\mathbf{k}
)\right)&&\left( 
\begin{array}{cc}
\bm{\epsilon}\cdot \hat{\mathbf{v}} & 0 \\ 
0 & -(\bm{\epsilon}\cdot \hat{\mathbf{v}})^{\ast }
\end{array}
\right) \left( 
\begin{array}{c}
u_{N}(\mathbf{k}) \\ 
v_{N}(\mathbf{k})
\end{array}
\right),
\end{eqnarray}
where $\hat{\mathbf{v}}=\nabla _{\mathbf{k}}\hat{H}_{0}(\mathbf{k})/\hbar $
is the velocity operator. In the tight-binding representation of the Sr$_{2}$
RuO$_{4}$ bands \cite{Annett2009}, the wave functions are $u_{N}(\mathbf{k}
)\equiv u_{N}^{m\sigma }(\mathbf{k})$ and $v_{N}(\mathbf{k})\equiv
v_{N}^{m\sigma }(\mathbf{k})$, where the orbital index $m$ runs over the
three Ru $4d$ orbitals $(d_{xz},d_{yz},d_{xy})$ and the index $\sigma $
represents electron spin. In this basis $\hat{H}_{0}(\mathbf{k})$ is the
$6\times 6$ tight binding Hamiltonian, including both on-site energies,
hopping integrals and, in general, spin orbit interactions.
Most of the calculations described below  have been performed 
for the set of parameters used
earlier \cite{Annett2003} in our modeling of strontium ruthenate with
non-zero out-of-plane inter-orbital interactions between  $d_{xz}$ and 
$d_{yz}$ orbitals. 

\begin{figure}[htb]
\includegraphics[width=15pc]{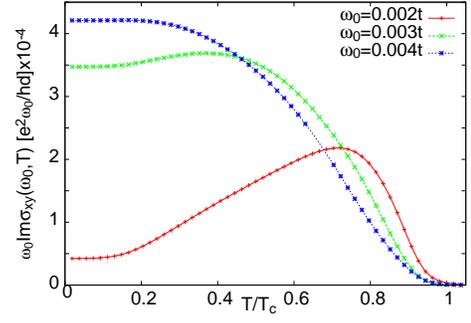}
\caption{The temperature dependence of the dichroic signal in the chiral
state calculated for a few values of the probing light frequencies. For this
particular set of interaction parameters $T_c=0.00135t$, which is slightly lower than 
$0.0015t$ corresponding to $T_c \approx 1.5K$.}
\label{dich-a-T}
\end{figure}

We start the discussion by showing in Fig. (\ref{dich-a-T}) 
the temperature dependence of the
imaginary part of the three dimensional Hall conductivity 
$Im\sigma_{xy}(T,\omega)$ calculated for a number of frequencies $\omega$.
Note, that the results have been shown in natural units for 3 dimensional
conductivity $i.e.$ ${\frac{e^2 }{hd}}$, where $e$ is the electron charge, $
h $ - Planck's constant and $d$ the c-axis lattice constant, $d=1.3nm$ in
strontium ruthenate. The energies are measured in units of $t$ - the
in-plane hoping parameter between $d_{xy}$ orbitals, which has been
estimated to be $t=0.08162~eV$.

\begin{figure}[htb]
\includegraphics[width=15pc]{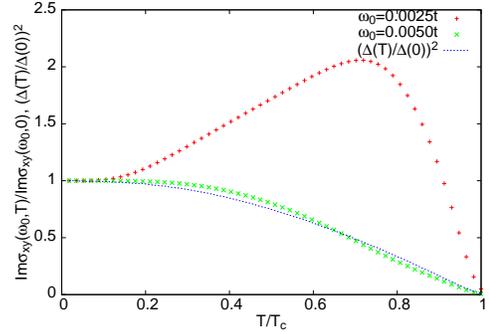} 
\caption{The temperature dependence of the $Im\protect\sigma_{xy}(\protect
\omega_0,T)$ normalized to its low temperature value in the chiral state for
two values of the light frequency: slightly below the zero temperature gap
value $\protect\omega_0=0.0025t$ and above it $\protect\omega_0=0.0050t$
compared to the normalized gap $({\frac{\Delta(T)}{\Delta(0)}})^2$. Note
the roughly quadratic dependence of the dichroic signal on the gap for the
probing frequency larger than the gap, and the strong departures from such a
behavior for low optical frequencies. }
\label{peak-a-T}
\end{figure}

In Fig. (\ref{dich-a-T}) the frequencies $\omega_0$ range from smaller than
the zero temperature energy gap $\Delta(0)\approx 0.0033t$ in the $d_{xz}$
and $d_{yz}$ orbital space, to larger than it. In the low frequency case a
coherence peak is observed, which is absent for higher optical frequencies. 
The temperature dependence of $Im\sigma_{xy}$ is
easily related to that of the superconducting gap in the large frequency
limit were it scales approximately as second power of the gap. The curves
normalized to their low temperature values, are shown in Fig. (\ref{peak-a-T}). 
It is worth noting that while the high frequency signal
scales roughly as the the square of normalized order parameter, the low
frequency results show strong deviations, which can be identified as a
coherence peak similar to the Hebel-Slichter \cite{hebel1957} peak observed
in NMR experiments on classic superconductors. This
coherence peak in the temperature dependence of the dichroic signal is not
apparent in the experiment \cite{Xia2006,Kapitulnik2009}, which was in the 
high frequency limit.  For this system the observation of the coherence peak would
require usage of light with low frequencies of the order $\omega_{cp}\approx
0.003 t= 0.245 meV$, $i.e.$ in the far infrared region of the spectrum.

The reflection coefficient $|r|$ and the polar Kerr angle $\theta_K$ are
given by the following equations \cite{Lutchyn2008,White-Geballe} 
\begin{eqnarray}
&& |r|=\frac{|n-1|}{|n+1|},  \label{r} \\
&& \theta_K=\frac{4\pi}{\omega}\,\mathrm{Im}\frac{\sigma_{xy}(\omega)}{
n\,(n^2-1)},  \label{theta_K}
\end{eqnarray}
where $n$ is the complex refraction coefficient. The polar Kerr angle 
(\ref{theta_K}) has been found \cite{Lutchyn2008} in the high frequency regime ($
\omega > \omega_{ab}$) to read 
\begin{equation}
\theta_K=\frac{4\pi\omega^2\,Im\sigma_{xy}(\omega)} {\sqrt{
\epsilon_\infty\omega^2-\omega_{ab}^2} \,
[(\epsilon_\infty-1)\omega^2-\omega_{ab}^2]},  \label{theta_K-high}
\end{equation}
and 
\begin{equation}
\theta_K=-\frac{4\pi\omega^2\,Re\sigma_{xy}(\omega)} {\sqrt{
\omega_{ab}^2-\epsilon_\infty\omega^2} \,
[(\epsilon_\infty-1)\omega^2-\omega_{ab}^2]},  \label{theta_K-low}
\end{equation}
for light frequencies smaller than in-plane plasma frequency $\omega_{ab}$.

\begin{figure}[htb]
\includegraphics[width=15pc]{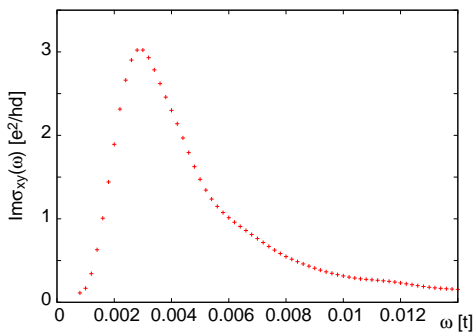}
\caption{The frequency dependence of $Im\protect\sigma_{xy}$ calculated for
the chiral state at low temperature.}
\label{dichr2afr2}
\end{figure}

The frequency dependence of the $Im\sigma_{xy}(\omega)$ is shown in the 
Fig.(\ref{dichr2afr2}) for low frequencies and temperature close to $0K$.

The approach we use here gives us an access to the elements $Im
\sigma_{xy}(\omega,T)$ and $Re\sigma_{xx}(\omega,T)$ of the conductivity
tensor. To calculate $Re\sigma_{xy}(\omega)$ needed to calculate $\theta_K$
in the frequency limit appropriate for experiments ($\Delta\ll\omega<
\omega_{ab}$) one has to perform Kramers-Kronig analysis \cite{bennett1965}.
To this end the full frequency dependence of the $Im\sigma_{xy}(\omega)$ is
needed. Assuming that at very high frequencies $\omega Im\sigma_{xy}(\omega)$
tends to a constant we obtain $Re\sigma_{xy}(\omega=0.8eV=9.8t)\approx 1.8
~ 10^{-6}$ in natural units $e^2/(hd)$. 
This number together with the approximation 
\begin{equation}
\theta_K=4\pi\frac{\omega^2} {\omega_{ab}^3}Re\sigma_{xy}(\omega),
\end{equation}
and the experimental value of plasma frequency $\omega_{ab}=4.5 eV\approx
55.1t$ gives $\theta_K\approx 200$ nrad, which is reasonably close to the experimental
value of order of $90$ nrad.

The dichroic signal we obtain from the  imaginary part of the Hall conductivity 
changes sign with chirality of the state $\sin k_x \pm i\sin k_y$ and, as expected,
equals exactly zero for non-chiral states. In the normal state the appearance 
of the dichroic signal  requires
both spin-orbit coupling and an external magnetic field breaking time reversal
symmetry \cite{Capelle1998}. 

In the present calculations the non-zero dichroic signal we obtain
for the chiral state of Sr$_2$RuO$_4$ can be shown to
arise from inter-orbital ($d_{xz}$, $d_{yz}$) Cooper pairs. The 
signal becomes zero if we remove
the pairing interaction for these inter-orbital pairs in our model, 
leaving only $d_{xy}$ orbital pairing on a single sheet of Fermi surface. 
Using single band models Lutchyn {\it et al.} \cite{Lutchyn2008,Lutchyn2009} 
and Goryo \cite{Goryo2008,Goryo2010} have found
a non-zero Kerr effect only by considering the scattering of carriers by impurities,
and therefore this is
an extrinsic Kerr effect. In a very clean system, like strontium ruthenate,
this third order impurity scattering might seem improbable to be solely
responsible for the measurements.
In a very recent paper Taylor and Kallin\cite{taylor2011} have also proposed a very similar 
theory for an intrinsic interband contribution to the Kerr effect in Sr$_2$RuO$_4$.

An  experimental test of our mechanism is possible because 
the temperature dependence of the Hall conductivity is not universal.
In our mechanism it
shows a coherence peak similar to that found by Hebel and Slichter in the
temperature dependence of nuclear relaxation time 1/T$_{1}$ as measured in
NMR. This prediction \cite{Capelle1998} can, in principle, be
tested experimentally by changing the frequency of the light.
This would allow the present mechanism to be compared to other possible
sources of dichoism, either arising from collective excitations 
\cite{Klemm1988,Hirschfeld1989,Yip1992} or high order impurity scattering 
\cite{Goryo2008,Goryo2010,Lutchyn2008,Lutchyn2009}.

Interestingly the presence of multiple bands around the Fermi energy 
occurs for many superconductors and in these systems the presence of at least a 
small inter-orbital/inter-band contribution to the pairing is very likely. Thus 
the mechanism which we propose may be operative not only in Sr$_2$RuO$_4$ but also in 
other systems, such as some high
temperature superconductors \cite{Kapitulnik2009}.

Finally it is of interest to recall that for normal sytems the
integral
\begin{equation}
\langle \mathrm{Im}[\sigma _{xy}(\omega )]\rangle \equiv \int_{0}^{\infty }
\mathrm{Im}[\sigma _{xy}(\omega )]d\omega .  \label{dsr}
\end{equation}
is related to a certain component of the orbital magnetization
$\overrightarrow{M}$ by the f-sum rule.
This was first derived by Oppeneer \cite{Oppeneer}
and further dicussed by Souza and Vanderbilt \cite{Souza2008}.
Clearly, if a similar relation held for superconductors it could lead to
new insights into the highly controversial question of what is the total
orbial momentum of a p-wave superconductor. Indeed using (\ref{dsr}) and following
the arguments of Souza and Vanderbilt we find
\begin{eqnarray}
\langle \mathrm{Im}[\sigma_{xy}^{z}(\omega )]\rangle &=& 
 \frac{\pi ^{2}e^{2}}{V} 
 \left( tr [\widehat{P}_{u,u}\widehat{r}\times \widehat{Q}_{u,u}\widehat{v}]   
 \right. \nonumber\\  
  && \left. -  tr [\widehat{P}_{v,v}\widehat{r}\times 
  \widehat{Q}_{v,v}\widehat{v}]\right)_{z}
 + \Sigma _{x,y}^{z} \label{eqntwelf} 
\end{eqnarray}
where $\widehat{r}$ and $\widehat{v}$ are the position 
and velocity operators, respectively,
and 
the particle and hole projection operators are defined as 
$\widehat{P}_{u,u}=\sum_{N}\left\vert u_{N^{\prime }}\right\rangle
f_{N}\left\langle u_{N}\right\vert $ and $\widehat{Q}_{u,u}=\sum_{N}
\left\vert u_{N^{\prime }}\right\rangle (1-f_{N})\left\langle
u_{N}\right\vert $, respectively, with $f_N$ the Fermi Dirac distribution of the 
 quasiparticle state of energy $E_N({\bf k})$. 
In Eq. (\ref{eqntwelf})
the contribution 
$\Sigma_{x,y}^{z}$ is given by
\begin{eqnarray}
\Sigma _{x,y} =\frac{\pi ^{2}e^{2}}{2V}\sum\limits_{N,N^{\prime
}} && \{f_{N}x_{N,N^{\prime }}^{u,u}\left( 1-f_{N^{\prime }}\right)
v_{y;N^{\prime }N}^{v,v}  \label{eqnthirteen} \nonumber \\
&&-f_{N}y_{N,N^{\prime }}^{v,v}\left( 1-f_{N^{\prime }}\right)
v_{x;N^{\prime }N}^{u,u}  \nonumber \\
&&-f_{N}x_{N,N^{\prime }}^{v,v}\left( 1-f_{N^{\prime }}\right)
v_{y;N^{\prime }N}^{u,u}  \nonumber \\
&& +f_{N}y_{N,N^{\prime }}^{u,u}\left( 1-f_{N^{\prime }}\right) 
v_{x;N^{\prime}N}^{v,v}\}  \label{eqsigma}
\end{eqnarray}
where, for brevity, 
$x_{N,N^{\prime }}^{u,u}\equiv \langle u_N \mid x \mid u_{N\prime}\rangle $,
$v_{y;N^{\prime }N}^{v,v}\equiv \langle v_{N\prime} \mid v_y \mid v_N\rangle $  etc..

The first term in Eq. (\ref{eqntwelf}) is a contribution to the total angular
momentum given by the particles and holes separately.  One may regard it as a
quasiparticle conribution to the orbital magnetization. 
Reassuringly, in the normal state it reduces to the
component of the orbital magnetization defined
 by Souza and Vanderbilt \cite{Souza2008} as 
$M_{RS}$. On the other hand, the second 
contribution in Eq. (\ref{eqntwelf}), $\Sigma _{x,y}^{z}$, 
as can be seen in Eq. (\ref{eqsigma}), involves products of both particle 
and hole amplitudes and
therefore can be regarded as the consequence of the order parameter, namely
the condensate. Further discussion of this very interesting f-sum rule will
be published elswhere \cite{gyorffy2011}. 

\begin{figure}[htb]
\includegraphics[width=15pc]{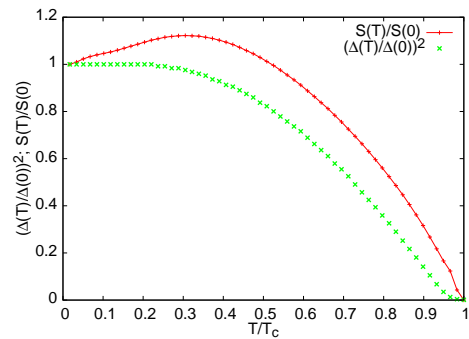}
\caption{The temperature dependence of  the f-sum rule 
$S(T)\equiv\langle \mathrm{Im}[\sigma_{xy}^{z}(\omega )]\rangle$
 normalized to its zero 
temperature value $S(0)$. This is compared with the 
square of normalized order parameter 
$(\Delta(T)/\Delta(0))^2$.}
\label{3sum-norm}
\end{figure}

Here we merely note that the f-sum rule for Sr$_2$RuO$_4$, shown in 
Fig. (\ref{3sum-norm}), also has a characteristic temperature dependence,
which can be compared with experiments and with other theories of 
orbital magnetization in the chiral pairing state.  For example
we can compare this temperature dependence with that which we 
calculated previously \cite{Annett2009} with the same tight-binding
Hamilonian and model gap equation for  Sr$_2$RuO$_4$ as discussed in this letter.  
We previously estimated that the orbital magnetization $M_{RS}^{z}$ in the chiral 
superconducting state  had a temperature dependence which fitted very
well with that of $\left( \frac{\Delta (T)}{\Delta (0)}\right) ^{2}$. 
It is clear from Fig. (\ref{3sum-norm}) that this gives 
a reasonable, but not perfect,
fit to the results obtained from the f-sum rule. 
The previous calculation \cite{Annett2009}
evaluated the magnetization in 
a theory which only included the first, quasiparticle, terms in 
Eq. (\ref{eqntwelf}). Thus we attribute the 
corresponding deviation in Fig. (\ref{3sum-norm})
to the contribution of the of the condensate terms, Eq. (\ref{eqsigma}).
This suggests that the mechanism of dichroism 
arising from inter-orbital pairing
discuused in this letter operates through both the quasi-particle extations
and the condensate to produce the total contribitions to the dichroic signal.

In conclusion, we predict the existence of an intrinsic dichroic signal in
systems with inter-orbital/inter-band Cooper pairs with chiral symmetry of
the order parameter. These calculations also suggest that a non-zero Hall
conductivity may also arise in other materials having intra-orbital order
parameters, with differing phases of the order parameters between the 
various orbitals. In this
case the inter-orbital/inter-band Josephson-like coupling is ultimately 
responsible for the effect.

\medskip This work has been partially supported by the Ministry of Science
and Higher Education grant No. N N202 2631 38. One of us (KIW) is grateful
to the stuff of MPI PKS in Dresden for the hospitality extended to him
during early stages of the present work


\begin{thebibliography}{99}
\bibitem{QZ10} Xiao-Liang Qi and Shou-Cheng Zhang,  	
Rev. Mod. Phys. {\bf 83}, 1057  (2011).

\bibitem{Mackenzie2003} A.P. Mackenzie and Y. Maeno, Rev. Mod. Phys. \textbf{%
75}, 657 (2003).

\bibitem{Luke1998} 
G.M. Luke, Y. Fudamoto, K.M. Kojima \textit{et al.} Nature \textbf{394} 558
(1998).

\bibitem{Kapitulnik2009} 
A. Kapitulnik, J. Xia, E. Schemm, \textit{et al.} New J. Phys. \textbf{11}
055060 (2009).




\bibitem{Klemm1988} R.A. Klemm,K. Scharnberg, D. Walker et al., Z. Phys. B \textbf{72} 139
(1988).

\bibitem{Hirschfeld1989} 
P.J. Hirschfeld, P. Wolfle, J.A. Sauls, D. Einzel and W.O. Putikka, Phys.
Rev. B \textbf{40} 6695 (1989).

\bibitem{Joynt1991} 
Q.P. Li and R. Joynt, Phys. Rev. B \textbf{44} 4720 (1991).

\bibitem{Yip1992} 
S.K. Yip and J.A. Sauls, J. Low. Temp. Phys. \textbf{86} 257 (1992).

\bibitem{Xia2006} 
Jing Xia, Yoshiteru Maeno, Peter T, Beyersdorf, et al., Phys. Rev. Lett. 
\textbf{97} 167002 (2006).



\bibitem{Hicks2010} 
C.W. Hicks, J.R. Kirtley, T.M. Lippman \textit{et al.} Phys. Rev. B \textbf{%
81} 214501 (2010).

\bibitem{Ashby2009} 
P.E.C. Ashby and C. Kallin, Phys. Rev. B \textbf{79} 224509 (209).

\bibitem{Kallin2008} 
C. Kallin and A.J. Berlinsky, 
J. Phys. Condens. Matter \textbf{21} 164210 (2009).

\bibitem{Yakovenko2007} 
V.M. Yakovenko, Phy. Rev. Lett. \textbf{98} 087003 (2007).

\bibitem{Mineev2008} 
V.P. Mineev, Phys. Rev. B \textbf{77} 180512 (2008).

\bibitem{Roy2008} 
R. Roy and C. Kallin, Phys. Rev. B \textbf{77} 174513 (2008).

\bibitem{Lutchyn2008} 
R.M. Lutchyn, P. Nagornykh and V.M. Yakovenko, Phys. Rev. B \textbf{77}
144516 (2008).

\bibitem{Goryo2008} 
J. Goryo, Phys. Rev. B \textbf{78} 060501 (2008)

\bibitem{Lutchyn2009} 
R.M. Lutchyn, P. Nagornykh and V.M. Yakovenko, Phys. Rev. B \textbf{80}
104508 (2009)

\bibitem{Goryo2010} 
J. Goryo, Mod. Phys. Lett. B \textbf{24} 2831 (2010).

\bibitem{Annett2002} J.F. Annett, G. Litak, B.L. Gyorffy and K.I.
Wysokinski, Phys. Rev. B \textbf{66}, 134514 (2002).

\bibitem{Annett2003} J.F. Annett, B.L. Gyorffy, G. Litak, and K.I.
Wysokinski, Eur. Phys. J. B \textbf{36}, 301 (2003).

\bibitem{Annett2006} J.F. Annett, G. Litak, B.L. Gyorffy and K.I.
Wysokinski, Phys. Rev. B \textbf{73}, 134501 (2006).

\bibitem{Annett2009} J.F. Annett, B.L. Gyorffy and K.I. Wysokinski, 
New J. Phys. \textbf{11} 055063 (2009).

\bibitem{Souza2008} 
I. Souza and D. Vanderbilt, Phys. Rev. B \textbf{77} 054438 (2008).

\bibitem{Raghu2010} 
S. Raghu, A. Kapitulnik and S.A. Kivelson, Phys. Rev. Lett. \textbf{105}
136401 (2010).

\bibitem{Capelle1997} 
K. Capelle, E.K.U. Gross and B.L. Gyorffy, Phys. Rev. Lett. \textbf{78} 3753
(1997).

\bibitem{Capelle1998} 
K. Capelle, E.K.U. Gross and B.L. Gyorffy, Phys. Rev. B \textbf{58} 473
(1998)

\bibitem{hebel1957} L C Hebel and C P Slichter, Phys. Rev. \textbf{107} 901
(1957).

\bibitem{Oppeneer} P.M. Oppeneer, J. Magn. Magn. Mat. {\bf 188} 275-285 (1998).


\bibitem{White-Geballe} R.M.~White and T.H.~Geballe, \emph{Long Range Order
in Solids} (Academic, New York, 1979), pp. 317, 321.

\bibitem{bennett1965} H.S. Bennett and E.A. Stern, Phys. Rev. \textbf{137},
A488 (1965).

\bibitem{taylor2011}  E. Taylor and C. Kallin arXiv:1111.447.

\bibitem{gyorffy2011} B.L. Gyorffy, unpublished
\end{thebibliography}
\end{document}